\documentstyle[12pt]{article} 

\newcommand{\be}{\begin{equation}}
\newcommand{\ee}{\end{equation}}
\newcommand{\bea}{\begin{eqnarray}}
\newcommand{\eea}{\end{eqnarray}}
\def \ppsl {p \kern-.45em{/}}
\def \lsim {\raisebox{-.7ex}{$\stackrel{\textstyle <}{\sim}\,$}}
\def \ct   {c_{\mbox{\footnotesize{w}}}}
\def \st   {s_{\mbox{\footnotesize{w}}}}
\def \xiw  {\xi_{\mbox{\footnotesize{w}}}}
\def \Mw   {M_{\mbox{\footnotesize{w}}}}
\def \Mz   {M_{\mbox{\footnotesize{z}}}}
\def \sov  {\overline{s}}

\begin{document}              

\begin{titlepage}

\hspace{11.2cm}	BNL~-~63280 

\hspace{11.2cm} MPI-PhT/96-51

\begin{center}
\vspace{0.5cm}
{\large\bf ANALYSIS OF THE \boldmath{$Z^0$} RESONANT AMPLITUDE \\
           IN GENERAL \boldmath{$R_\xi$} GAUGES}

\vspace{1cm}
\renewcommand{\thefootnote}{\fnsymbol{footnote}}
{\bf	Massimo Passera$^a$\footnote{E-mail address: 
	passera@mafalda.physics.nyu.edu} ~and~ Alberto 
	Sirlin$^b$\footnote{Permanent address: Dept. of Physics, New York 
	University, 4 Washington Place, New York, NY 10003, USA. 
	E-mail address: sirlin@mafalda.physics.nyu.edu}}

\vspace{1cm}
{\it	
	$^a$ Physics Department, Brookhaven National Laboratory, \\
	Upton, NY 11973-5000, USA

	$^b$ Max-Planck-Institut f\"{u}r Physik, \\
	F\"{o}hringer Ring 6, 80805, M\"{u}nchen, Germany}

\vspace{2cm}
{\large\bf Abstract}
\end{center}

\noindent
The $Z^0$ resonant amplitude is discussed in general $R_\xi$ gauges.
When the original on-shell definition of the $Z^0$ mass $M$ is employed,
a gauge dependence of $M$ emerges in the next-to-leading approximation which,
although small, is of the same magnitude as the current experimental error.
In the following order of expansion, these unphysical effects are unbounded.
The gauge dependence of $M$ disappears when modified, previously proposed 
definitions of mass or self-energies, are used. The relevance of these
considerations to the concept of the mass of unstable particles is pointed 
out.


\end{titlepage}


	Ever since LEP (Large Electron Positron collider)
commenced operations, major efforts have been devoted
by experimentalists and theorists to study the $Z^0$ line shape. Indeed,
this important observable leads to the determination of some of the most 
fundamental parameters in electroweak physics, namely the $Z^0$ mass, its 
width and the cross sections at the peak. Of primary interest in the 
theoretical side of these studies is the structure of the transverse part of 
the dressed $Z^0$ propagator
\be
	D(s)= \left( s- M_0^2 - Re A(s) -i Im A(s) \right)^{-1}, 
\label{eq:prop}
\ee
where $s=q^2$ is the squared four-momentum transfer, $M_0$ is the bare mass 
and $A(s)$ is the conventional $Z^0$ self-energy, which we have split into 
its real and imaginary components.

In the original formulation of the on-shell method of renormalization 
\cite{Si80}, the physical mass $M$ was related to the bare mass by means of
the expression
\be
	M^2 = M_0^2 + Re A(M^2),
\label{eq:M80}
\ee
so that the mass counterterm $\delta M^2 = M^2 - M_0^2$ was identified with 
$Re A(M^2)$. Recalling that counterterms in field theory are real, 
Eq.~(\ref{eq:M80}) is the simplest generalization to unstable particles of the
well known expression $m_e= m_e^0 + \Sigma(\ppsl = m_e)$ for the mass 
renormalization of the electron in QED. In particular, Ref.~\cite{Si80}
contains a rather detailed discussion of the gauge invariance of the 
resulting one-loop corrections investigated in that work. Since 1980 
Eq.~(\ref{eq:M80}) has been adopted by many theoretical physicists and, 
in fact, several standard analyses of the $Z^0$ line shape are based on such 
definition (see, for example, Refs.~\cite{BuH88,BCMS}). However, it was later
pointed out that, in spite of its simplicity and usefulness in the evaluation 
of the one-loop corrections, Eq.~(\ref{eq:M80}) contains theoretical 
limitations
in higher orders of perturbation theory. Using special arguments, it was 
concluded in Refs.~\cite{Si91a} and \cite{Si91b} that the use of 
Eq.~(\ref{eq:M80}) would generate gauge-dependent electroweak corrections in 
$O(g^6)$ and, in a restricted class of $R_\xi$ gauges, even in $O(g^4)$.
Since the pioneering work of W.~Wetzel \cite{Wet}, effects of $O(g^4)$ are 
routinely incorporated in the analysis of $D^{-1}(s)$. In fact, in the 
resonance region $|s-M^2|$ \lsim $M \Gamma$, and therefore the leading 
contributions in $D^{-1}(s)$ are of $O(g^2)$. Thus, in the next-to-leading
order (NLO) approximation employed by theorists to study the resonant 
amplitude, one must retain all contributions of $O(g^4)$ in $D^{-1}(s)$. 
Contributions of $O(g^6)$ in $D^{-1}(s)$ should be taken into account when 
one expands $D(s)$ around the resonance in order to obtain the corresponding 
contribution to the non-resonant amplitude. 
So far the gauge dependence induced by 
Eq.~(\ref{eq:M80}) has not been explicitly demonstrated in the analysis of the
line shape. Fortunately, it was also anticipated in Refs.~\cite{Si91a,Si91b}
that the $O(g^4)$ gauge dependence is numerically bounded, so that the effect
in this order is expected to be small.

In this paper we reexamine the electroweak corrections to Eq.~(\ref{eq:prop})
in the resonant region. Unlike previous studies, we work in the framework of
general $R_\xi$ gauges. Our aim is three-fold: {\it i}) to explicitly 
show that, as anticipated in Refs.~\cite{Si91a,Si91b}, a gauge dependence 
emerges when Eq.~(\ref{eq:M80}) is employed, 
{\it ii}) to evaluate its magnitude 
in $O(g^4)$, {\it iii}) to explicitly show that the gauge dependence 
disappears when previously proposed, modified definitions of $M$ or $A(s)$, 
are employed.

We first discuss Eq.~(\ref{eq:prop}) in the NLO approximation. Inserting 
Eq.~(\ref{eq:M80}) into Eq.~(\ref{eq:prop}) and recalling that in the 
resonance region $s-M^2$ is $O(g^2)$, we have 
\bea
	D^{-1}(s) & = & (s-M^2) \left[ 1- Re A'(M^2) \right]
					 -i Im A(M^2)   \nonumber \\
                  &   & -i Im A'(M^2)(s-M^2) + O(g^6).
\label{eq:iprop}
\eea
For $Im A(M^2)$ we can employ the unitarity relation 
\be
	-i Im A(M^2) = i M \Gamma \left[1 - Re A'(M^2)\right],
\label{eq:uni}
\ee
where $\Gamma$ is the radiatively corrected width.
As $s-M^2 =O(g^2)$ it is sufficient to evaluate $Im A'(M^2)$ in the last 
term to one-loop order. Calling $A_f$ and $A_b$ the fermionic and bosonic 
contributions, we have
\be
	Im A'_f(M^2) = -\Gamma/M - (3/4 \pi \sqrt{2})G_{\mu} m_b^2.
\label{eq:sca}
\ee
The last term, which is very small, arises from the leading violation to 
the scaling behavior $Im A_f(s) \sim s$, due to the finiteness of the $b$
quark mass $m_b$. Its magnitude can be gleaned, for example, from Eq.~(25)
of Ref.~\cite{DS91} or from Ref.~\cite{BCMS}. To evaluate $Im A'_b(M^2)$
in the $R_\xi$ gauges, we need the expression of the one-loop self-energies
in such a general framework. They are given in compact form in 
Ref.~\cite{DS92}.
In the analysis we restrict ourselves to ${\xi_{\mbox{\footnotesize{w}}}}
\geq 0$, where ${\xi_{\mbox{\footnotesize{w}}}}$ is the $W$-gauge parameter.
Negative values of ${\xi_{\mbox{\footnotesize{w}}}}$ are not tenable
for at least two reasons:
{\it i)} as it is well-known, the euclidean path
integral representation of the generating functional does not exist in this
case (i.e. it does not converge);
{\it ii)} for ${\xi_{\mbox{\footnotesize{w}}}}<0$ the unphysical scalar
mass $m_{\phi^{\pm}} $ becomes imaginary and the $W^\pm$ and
$\phi^\pm$ propagators develop pathological singularities at space-like
values of $k^2$. Taking this restriction into account and evaluating the
corresponding imaginary parts through $O(g^2)$, we find
\bea
	Im A'_b(M^2) & = & -\theta [(4\ct^2)^{-1} - \xiw] 
			(\alpha/24 \ct^2 \st^2)
			(1-4 \ct^2 \xiw)^{3/2}  \nonumber  \\
		     & + &  \theta [(\ct^{-1}-1)^2 -\xiw]
			(\alpha/12 \ct^2)
			(a^2 -4\ct^2)^{1/2}(a^2 +8\ct^2),
\label{eq:im}
\eea
where $c_{\mbox{\footnotesize{w}}} \equiv
M_{\mbox{\footnotesize{w}}}/M_{\mbox{\footnotesize{z}}}$ and
$a \equiv 1 + \ct^2(1-\xiw)$. To understand the origin of these 
gauge-dependent contributions, we recall that the relevant unphysical fields
have masses $\Mw \xiw^{1/2}$. When $\xiw$ is sufficiently small, $A_b(s)$
develops imaginary parts in the neighborhood of $s=M^2$, which contribute 
gauge-dependent contributions to $Im A'_b(M^2)$. We expect such contributions 
when $\Mz \geq 2 \Mw \xiw^{1/2}$, i.e. $\xiw \leq (4\ct^2)^{-1}$ and when
$\Mz \geq \Mw(1+\xiw^{1/2})$, i.e. $\xiw \leq (\ct^{-1}-1)^2$. They 
correspond to the two terms in the r.h.s. of Eq.~(\ref{eq:im}). Inserting
Eqs.~(\ref{eq:uni},~\ref{eq:sca},~\ref{eq:im}) in Eq.~(\ref{eq:iprop}), we 
rewrite this expression in the form
\be
	D^{-1}(s) = \left(1- Re A'(M^2) - i \Phi\right)
		    \left(s- \widetilde{M}^2 + i s \widetilde{\Gamma}/
					\widetilde{M}\right) + O(g^6),
\label{eq:iprop2}
\ee
\bea
	\Phi  &\equiv& Im A'_b(M^2) - (3/4 \pi \sqrt{2})G_{\mu} m_b^2,
					\label{eq:Phi} \\
	\widetilde{M}^2  &\equiv& M^2 (1+\Phi \Gamma/M), 
					\label{eq:Mtil} \\
	\widetilde{\Gamma}/\widetilde{M} & \equiv & \Gamma/M.
\eea
The gauge dependence of $Re A'(M^2)+i \Phi$ in Eq.~(\ref{eq:iprop2}) cancels
against corresponding contributions in the vertex functions. The very small
scaling violation in $i \Phi$ is shifted to higher orders when evaluating
$|D(s)|^2$ ~\cite{BCMS}. The amplitude $(s- \widetilde{M}^2 + i s 
\widetilde{\Gamma}/{\widetilde{M}})$ has the characteristic $s$-dependent 
Breit-Wigner form employed in the LEP analysis. It is clear, however, that
what LEP measures is $\widetilde{M}$ rather than M. As $\Phi$ in 
Eq.~(\ref{eq:Phi}) contains the gauge-dependent contribution $Im A'_b(M^2)$,
it follows from Eq.~(\ref{eq:Mtil}) that $M$ is gauge-dependent. To evaluate 
the magnitude of the gauge dependence we note that the maximum value of
$\left|Im A'_b(M^2)\right|$ in Eq.~(\ref{eq:im}) occurs at 
$\xiw = (\ct^{-1}-1)^2$ (the threshold of the second contribution), in which
case $Im A'_b(M^2)= -1.6 \times 10^{-3}$. On the other hand, 
$Im A'_b(M^2)=0$ for $\xiw \geq (4\ct^2)^{-1}$. Thus the maximum shift in 
$M$ due to the gauge-dependence is 
$|\delta M| = 1.6 \times 10^{-3} \Gamma/2 = 2$ MeV. Although this is a small
effect, it is of the same magnitude as the current experimental error
$\Delta M = \pm 2.2$ MeV \cite{LS}.

In order to circumvent the gauge dependence generated by Eq.~(\ref{eq:M80})
in higher orders, an alternative definition for the $Z^0$ mass was proposed 
in Refs.~\cite{Si91a,Si91b} which in $O(g^4)$ differs from Eq.~(\ref{eq:M80})
in very small but gauge-dependent terms. Calling 
$\sov = m_2^2 -i m_2 \Gamma_2$ the complex-valued position of the pole
in $D(s)$, so that $\sov-M_0^2-A(\sov)=0$, the physical mass and width 
were identified with 
\be 
	m_1^2 = m_2^2 + \Gamma_2^2, \qquad 
	\Gamma_1/m_1 = \Gamma_2/m_2.
\label{eq:m12}
\ee
Noting that $m_2^2=M_0^2+Re A(\sov)$, Eq.~(\ref{eq:m12}) corresponds to a 
mass counterterm 
\be
	\delta m_1^2 = m_1^2 -M_0^2 = Re A(\sov) + \Gamma_2^2.
\label{eq:delm1}
\ee
Through $O(g^4)$ this becomes 
$\delta m_1^2 = Re A(m_1^2)+ \Phi m_1 \Gamma_1 + O(g^6)$. The additional 
$O(g^4)$ term $\Phi m_1 \Gamma_1$ has important consequences. Repeating
the previous analysis one readily finds 
\be
	D^{-1}(s) = \left(1- Re A'(m_1^2) - i \Phi\right)
		    \left(s- m_1^2 + i s \Gamma_1/m_1\right) + O(g^6).
\label{eq:iprop3}
\ee
Again, the gauge dependence in $Re A'(m_1^2)+ i \Phi$ cancels against 
corresponding contributions in the vertex parts, and the scale violation 
in $i \Phi$ is promoted to higher orders in $|D(s)|^2$. It is important 
to note that, in contrast with Eq.~(\ref{eq:iprop2}), $m_1$ in the resonant 
factor is not modified. As a consequence, $m_1$ can be directly identified 
with the mass measured at LEP. In fact, there is an alternative and more 
elegant way for deriving Eq.~(\ref{eq:iprop3}), already outlined in 
Ref.~\cite{Si91a}. Recalling that $\sov$ is the position of the pole we
can write $D(s)= (s-\sov -A(s)+A(\sov))^{-1}$, the resonant part of which is 
$D^{res}(s)= (s-\sov)^{-1}(1-A'(\sov))^{-1}$. Multiplying and dividing by
$1+ i\Gamma_2/m_2$ and neglecting terms of higher order in the cofactor
$(1+ i\Gamma_2/m_2)/(1-A'(\sov))$ one obtains once more Eq.~(\ref{eq:iprop3}).
This shows that the $s$-dependent Breit-Wigner form automatically emerges
when the large imaginary part $-i Im A'(m_2^2)$ in the pole's residue is
absorbed into the resonant factor.

In order to obtain the leading non-resonant contribution, one expands
$D^{-1}(s)$ to the next order in $s-M^2$. If Eq.~(\ref{eq:M80}) is employed
we have 
\be
	D^{-1}(s)= s-M^2 - i Im A(M^2) - (s-M^2)A'(M^2) 
		- (s-M^2)^2 A''(M^2)/2 + O(g^8).
\label{eq:iprop4}
\ee
To simplify the analysis we now consider the class of gauges
$\xiw \geq (4\ct^2)^{-1}$, within which the previous calculation was
gauge independent, and further neglect the small scaling violation  in 
the one-loop amplitude $Im A_f(s)$. In this case we can set 
$Im A''(M^2)=0$ in Eq.~(\ref{eq:iprop4}). For $Im A(M^2)$ we again employ 
the unitarity relation (Eq.~(\ref{eq:uni})). After some elementary algebra, 
Eq.~(\ref{eq:iprop4}) becomes
\bea
	D^{-1}(s) &=&
	\left[\frac{1-A'(M^2 -iM\Gamma)}{1+i \Gamma/M}\right]
	\left(s- \widehat{M}^2 + i s 
	\frac{\widehat{\Gamma}}{\widehat{M}} \right) \times \nonumber \\
	&\times & \left[1-Re A''(M^2) \left(s- \widehat{M}^2 + i s 
	\widehat{\Gamma}/\widehat{M}\right)\!/2 \right] + O(g^8) ,
\label{eq:iprop5}
\eea
\be
	\widehat{M}^2 = M^2 -\delta, \qquad \qquad \qquad \qquad
	\widehat{\Gamma}/\widehat{M}= \Gamma/M,
\label{eq:Mhat}		
\ee
\be
	\delta = -Im A'(M^2)M\Gamma\left(1+Re A'(M^2)\right) -\Gamma^2 +
	Re A''(M^2)M^2\Gamma^2/2.
\label{eq:del}
\ee
In the first term of Eq.~(\ref{eq:del}), $Im A'(M^2)$ is evaluated through 
$O(g^4)$. In $O(g^2)$, $-Im A'(M^2)$ equals $\Gamma/M$ 
(cf.~Eq.~(\ref{eq:sca})); 
thus the $O(g^4)$ terms cancel in Eq.~(\ref{eq:del}) and $\delta$ is $O(g^6)$.
This quantity $\delta$ already occurred in the arguments of 
Refs.~\cite{Si91a,Si91b}, where it was pointed out that it is afflicted by 
an unbounded gauge dependence (i.e. it diverges in the unitary gauge). As the
LEP measurement can be identified with $\widehat{M}^2$, it follows from
Eq.~(\ref{eq:Mhat}) that $M$, defined in Eq.~(\ref{eq:M80}), is also afflicted
by an unbounded gauge dependence in $O(g^6)$. Once more the gauge dependence 
disappears if the physical mass is identified with Eq.~(\ref{eq:m12}).
In fact, expanding Eq.~(\ref{eq:delm1}) in the range $\xiw \geq (4\ct^2)^{-1}$,
one has $\delta m_1^2 = Re A(m_1^2) -\delta +O(g^8)$. The additional term 
cancels the gauge dependence in the resonant factor of
Eq.~(\ref{eq:iprop5}) 
and $m_1$ can be identified with the observed mass $\widehat{M}$.

Aside from Refs.~\cite{Si91a,Si91b}, a number of authors have advocated the 
idea of defining the $Z^0$ mass and width in terms of $m_2$ and $\Gamma_2$
\cite{var}. All such proposals should also lead to correct, gauge invariant 
answers. One significant phenomenological difference is that the definitions
in Eq.~(\ref{eq:m12}) lead to the $s$-dependent Breit-Wigner resonance 
employed in the LEP analysis, so that $m_1$ and $\Gamma_1$ can be 
identified with the LEP measurements. The other proposals, instead, differ 
numerically from such determinations by amounts much larger than the 
experimental error.

An alternative procedure to define a gauge invariant mass is to employ a 
gauge invariant self-energy in Eqs.~(\ref{eq:prop},~\ref{eq:M80}), instead of 
the conventional amplitude. The possibility of using the Pinch Technique 
(PT) self-energy was suggested in Ref.~\cite{DS-PT}. Recently, there has
been significant progress in the construction of the PT self-energies
in higher orders \cite{PP,BBBBW,KS}. In the expansion analogous to 
Eq.~(\ref{eq:iprop}) the conventional self-energy $A$ is replaced by its PT
counterpart $\widehat{A}$ which is $\xi$-independent and, moreover, 
$Im \widehat{A}'_b(M^2)=0$. Thus, the gauge dependence does not arise and
one obtains \cite{KS} an expression analogous to Eq.~(\ref{eq:iprop2}), 
with $Re A'(M^2) \rightarrow Re \widehat{A}'(M^2)$ and 
$\Phi \rightarrow \widehat{\Phi}= -(3/4 \pi \sqrt{2})G_{\mu} m_b^2$, a very 
small and gauge independent scale-breaking term. We have explicitly shown 
that the two methods described above eliminate the gauge dependence in 
the NLO approximation. However, the approach based on Eq.~(\ref{eq:m12}) 
also cancels the scale-breaking correction to the mass in 
Eqs.~(\ref{eq:Phi},~\ref{eq:Mtil}). 
In the $Z^0$ case this effect is extremely small,
a shift $\approx 0.05$ MeV in $m_1$. However, it may be more significant in 
other cases. Furthermore, in the PT approach the gauge independence of
$D^{-1}(s)$ in the resonance region has so far been demonstrated only
through $O(g^4)$ \cite{KS}. On the other hand, the PT analysis can be easily 
recast in a form very similar to Eq.~(\ref{eq:iprop3}), which conforms 
with the 
mass definition in Eq.~(\ref{eq:m12}) rather than Eq.~(\ref{eq:M80}). 
It suffices
to note that the PT result in any gauge can be expressed as \cite{KS}
\be
	D^{-1}(s) = (s-\sov)[1-\widehat{A}'(\sov)] + O(g^6),
\ee
which is explicitly $\xi$-independent and shows that the pole position is not 
displaced. Following the steps outlined after Eq.~(\ref{eq:iprop3}), one
obtains an expression analogous to that equation with 
$Re A'(m_1^2) \rightarrow Re \widehat{A}'(m_1^2)$ and 
$\Phi \rightarrow \widehat{\Phi}$. The difference with Eq.~(\ref{eq:iprop3})
is that the first factor (as well as the vertex parts) are now separately
$\xi$-independent. In other words, as far as it is presently known, in the
PT approach one can define the mass either from the expression analogous to 
Eq.~(\ref{eq:M80}) (with $A \rightarrow \widehat{A}$) or from 
Eq.~(\ref{eq:m12}).

Previous detailed studies of the $Z^0$ resonant amplitude have not uncovered
the gauge dependence found in the present analysis when Eq.~(\ref{eq:M80})
is employed. The reason is that the gauge-dependent contribution involving
$Im A'_b(M^2)$ is routinely disregarded, a procedure that is correct in the
subclass of gauges $\xiw \geq (4\ct^2)^{-1}$ (this includes the frequently
employed 't~Hooft-Feynman and unitary gauges). However, in the range
$\xiw \leq (4\ct^2)^{-1}$ (this includes the Landau gauge) one must consider 
such terms. Similarly, it appears that the contribution $\delta$, which
afflicts all gauges and is moreover unbounded, has also not been considered.
Ref.~\cite{Si91b} did detect these effects by sitting exactly at the
resonance and using indirect arguments. However, to make contact with 
experiment one must consider the full resonance region and demonstrate,
as shown in the present paper, how the gauge dependence arises in the 
analysis of the line shape. It is likely that the $O(g^4)$ gauge-dependent
effects discussed above are larger in the case of other unstable particles,
such as $W$ and $H$. In particular, in the $W$ case 
$Im A'^{\;\mbox{\footnotesize{w}}}_b(\Mw^2) \neq 0$ for $\xiw<1$, so that 
the gauge dependence induced by the definition analogous to Eq.~(\ref{eq:M80})
arises just below the 't~Hooft-Feynman gauge. These observations are relevant 
to elucidate the concept of mass for unstable particles, at least in the
context of gauge theories. Extrapolating the lessons learned in the $Z^0$
case to other particles, one is lead to the conclusion that such concept
should be based on the parameters that define the complex-valued position 
of the pole, as for example in Eq.~(\ref{eq:m12}), or on gauge-independent
self-energies, as in the discussions in the PT framework. In the first case,
Eq.~(\ref{eq:delm1}) gives the relevant mass counterterm in compact form.
Meanwhile, in cases in which $\Gamma$ is perturbatively small 
($\Gamma \ll M$), Eq.~(\ref{eq:M80}) remains a very useful approximation
that can be applied at the one-loop level and, in an important but 
restricted class of $R_{\xi}$ gauges, in $O(g^4)$.

\vspace{1.2cm}
\noindent
{\Large\bf Acknowledgments}
\vspace{0.6cm}

We are indebted to P.~Gambino, W.~Hollik, B.~Kniehl, M.~Porrati, M.~Schaden,
and D.~Zwanziger for very useful discussions.
One of us (A.~S.) would like to thank the Max-Planck-Institut f\"{u}r Physik
at Munich for its kind hospitality in the Summer of 1996. M.~P. would like to
express his gratitude to the Physics Department at the Brookhaven National 
Laboratory for its warm hospitality. This research was supported in part 
by the National Science Foundation under grant No. PHY-9313781, and the 
U.S. Department of Energy under Contract No. DE-AC02-76-CH00016.



\end{document}